\documentclass[runningheads]{llncs}
\usepackage[T1]{fontenc}
\usepackage{graphicx}
\usepackage{xcolor}

\begin{document}
\title{Denoising Diffusion Models for Inpainting of Healthy Brain Tissue}
%
%
\author{Alicia Durrer\orcidID{0009-0007-8970-909X} \and
Philippe C. Cattin\orcidID{0000-0001-8785-2713}\and
Julia Wolleb\orcidID{0000-0003-4087-5920}}
\authorrunning{A. Durrer et al.}
%
\institute{Center for medical Image Analysis and Navigation, Department of Biomedical Engineering, University of Basel, Allschwil, Switzerland}
\maketitle              
\begin{abstract}
This paper is a contribution to the "BraTS 2023 Local Synthesis of Healthy Brain Tissue via Inpainting Challenge". The task of this challenge is to transform tumor tissue into healthy tissue in brain magnetic resonance (MR) images. This idea originates from the problem that MR images can be evaluated using automatic processing tools, however, many of these tools are optimized for the analysis of healthy tissue. By solving the given inpainting task, we enable the automatic analysis of images featuring lesions, and further downstream tasks.
\newline Our approach builds on denoising diffusion probabilistic models. We use a 2D model that is trained using slices in which healthy tissue was cropped out and is learned to be inpainted again. This allows us to use the ground truth healthy tissue during training. In the sampling stage, we replace the slices containing diseased tissue in the original 3D volume with the slices containing the healthy tissue inpainting. With our approach, we achieve the second place in the challenge. On the test set our model achieves a mean SSIM of $0.8271$, a PSNR of $20.4949$ and a MSE of $0.0115$. In future we plan to extend our 2D model to a 3D model, allowing to inpaint the region of interest as a whole without losing context information of neighboring slices.  

\end{abstract}
\section{Introduction}

\begin{figure}
\includegraphics[width=\textwidth]{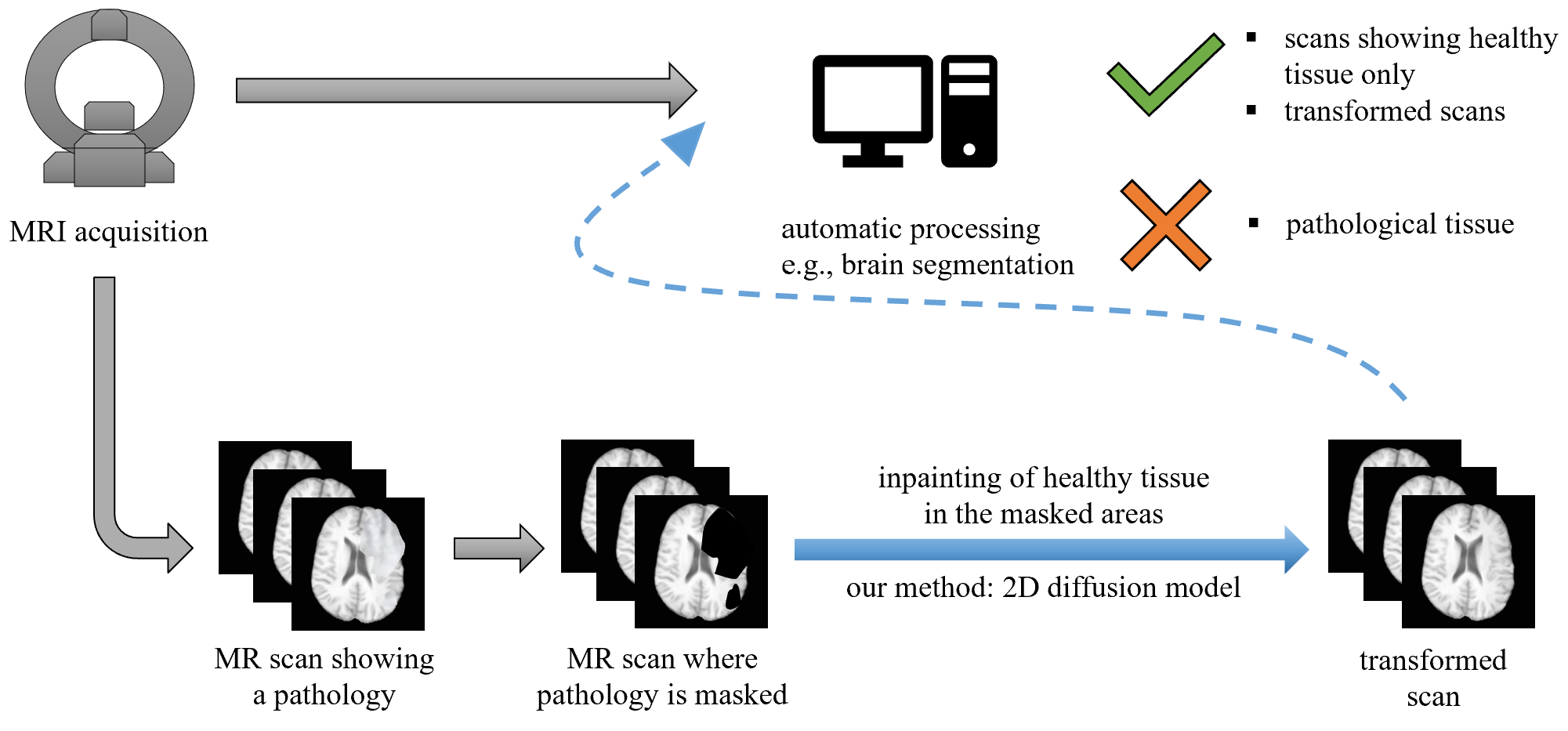}
\caption{Illustration of the problem. MR scans not showing pathologies can be fed directly into automatic processing tools. Often, scans containing pathologies need to be modified first. The task of this challenge is to replace pathological brain tissue with healthy brain tissue. The suggested inpainting requires a masked scan, whereby the masked region will be filled by an inpainting process, for which we suggest a DDPM-based model, visualized by the bold blue arrow. The output, denoted as the transformed scan, then shows a healthy scan. The transformed scan can then be fed into automatic processing tools for further analysis, denoted by the dashed blue arrow.} \label{fig1}
\end{figure}

Magnetic Resonance (MR) images of the brain and the subsequent automatic processing of these images are essential for monitoring pathologies. However, many automatic processing tools are built to e.g., register or segment only healthy tissue and are not as reliable for processing diseased brain regions~\cite{jenkinson2012fsl,farazi2016inpainting,almansour2021inpaint}. As illustrated in Figure~\ref{fig1}, it is therefore required to transform diseased tissue into healthy tissue to benefit from these automatic processing tools. This paper is a contribution to the "BraTS \cite{menze2014brats,bakas2017radiomic,bakas2018identifying,baid2021rsna} 2023 Local Synthesis of Healthy Brain Tissue via Inpainting Challenge"~\cite{kofler2023inpainting} that focuses on synthesizing healthy brain tissue in brains affected by tumors by transforming tumor tissue into healthy tissue. The challenge organizers provide T1-scans from the multi-modal BraTS 2022 segmentation challenge \cite{menze2014brats,bakas2017radiomic,bakas2018identifying,karargryris2023federated}. In these scans, the areas affected by glioma, tumors originating from glial cells in the brain or spinal cord~\cite{mamelak2007glioma}, need to be inpainted in three dimensions with healthy brain tissue.
For this task, we propose a method based on denoising diffusion probabilistic models (DDPMs) \cite{ho2020diffusion,saharia2022palette,wolleb2022diffusion}. We build on a method we previously applied to the task of contrast harmonization between different MR scanners~\cite{durrer2023diffusion}. 

\subsection{BraTS challenge}
BraTS \cite{menze2014brats,bakas2017radiomic,bakas2018identifying,baid2021rsna}, short for "International Brain Tumor Segmentation", provides challenges within the International Conference for Medical Image Computing and Computer-Assisted Intervention (MICCAI). The here discussed challenge "Local Synthesis of Healthy Brain Tissue via Inpainting"\cite{kofler2023inpainting}, is one of several challenges offered in the BraTS 2023 cluster of challenges.

\subsection{Contribution}
Based on our existing image-to-image translation method~\cite{durrer2023diffusion}, we train a slice-wise 2D DDPM to solve this inpainting task.

\section{Method}

\subsection{Denoising Diffusion Probabilistic Models}
DDPMs are generative models consisting of an iterative noising process $q$ and a learned denoising process $p_{\theta}$. Gaussian noise is added to an image $x$ for $T$ time steps $t$ during the noising process $q$. Each resulting image ${x_0, x_1, ..., x_T}$ contains a higher amount of noise than the previous image, as the noise level increases with each time step $t$ until it reaches a maximum at $t=T$. The equation

\begin{equation}
q(x_{t}|x_{t-1}):=\mathcal{N}(x_{t};\sqrt{1-\beta _{t}}x_{t-1},\beta _{t}\mathbf{I}),
\label{eqn1}
\end{equation}
describes the forward noising process $q$, whereby $\mathbf{I}$ is the identity matrix and $\beta_{1},...,\beta_{T}$ are the forward process variances. Defining $\alpha _{t}:=1-\beta _{t}$, $\overline{\alpha}_{t}:=\prod_{s=1}^t \alpha _{s}$ and applying the reparameterization trick, $x_t$ can be written as
\begin{equation}
x_{t}=\sqrt[]{\overline{\alpha} _{t}}x_{0}+\sqrt[]{1-\overline{\alpha} _{t}}\epsilon, \quad \mbox{with } \epsilon \sim \mathcal{N}(0,\mathbf{I}).
\label{eqn2}
\end{equation}
During the denoising process $p_{\theta}$, the goal is to reverse the forward process. Therefore, we want to predict $x_{t-1}$ from $x_t$ for $t \in \{T,...,1\}$, whereby $p_{\theta}$ is defined as a normal distribution with mean $\mu_{\theta}$ and variance $\Sigma_\theta$. The aim is to determine $\mu_{\theta}$ and $\Sigma_\theta$ such that $p_{\theta}$ and $q$ match. The denoising process $p_{\theta}$ can be written as
\begin{equation}
p_{\theta}(x_{t-1}\vert x_t):= \mathcal{N}\bigl(x_{t-1};\mu_{\theta}(x_t, t), \Sigma_\theta(x_t,t)\bigr).
\label{eqn3}
\end{equation}
By learning $\mu_{\theta}$, a model $\epsilon_{\theta}$ can be trained to denoise an image step by step. In \cite{ho2020diffusion} it was shown that $\Sigma_\theta$ can be fixed to $\sigma_{t}^2\mathbf{I}$ and does not need to be learned. The model $\epsilon_{\theta}$ is a U-Net~\cite{ronneberger2015unet} that predicts the noise pattern $\epsilon_{\theta }(x_{t},t)$ at time step $t$ to generate a slightly denoised image $x_{t-1}$.
In their original form, DDPMs are generative models that can sample images starting from random noise. 
This sampling process is given through:  
\begin{equation}
x_{t-1}=\frac{1}{\sqrt[]{\alpha _{t}}}\left(x_{t}-\frac{1-\alpha _{t}}{\sqrt[]{1-\overline{\alpha }_{t}}}\epsilon _{\theta }(x_{t},t)\right)+\sigma_{t}\mathbf{z}, \quad \mbox{with } \mathbf{z} \sim \mathcal{N}(0,\mathbf{I}),
\label{eqn4}
\end{equation}
whereby we start from $x_T$  $\sim \mathcal{N}(0,\mathbf{I})$ and apply Equation~\ref{eqn4} for all time steps $t \in \{T,...,1\}$. The final prediction is $x_0$. Section 2.2 describes how this well-known method is modified for inpainting.

\subsection{Inpainting of masked region}
\begin{figure}
\includegraphics[width=\textwidth]{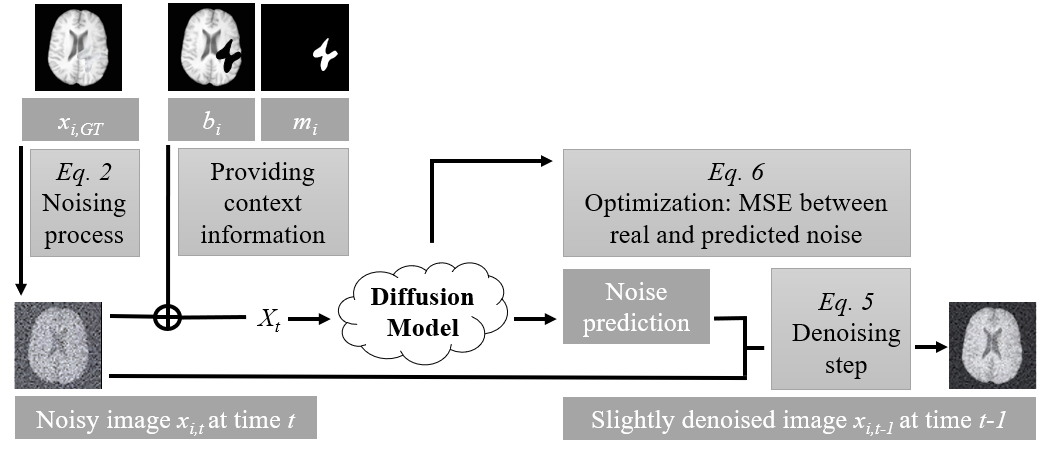}
\caption{Overview of the training process. Context information is provided through the concatenation of input $b_i$ and mask $m_i$ with the noisy image $x_{i,t}$, which originated from the ground truth $x_0$ = $x_{i,GT}$ using Equation~\ref{eqn2}. The concatenated image $X_t$ is used by the diffusion model to predict a slightly denoised image $x_{i,t-1}$ from $x_{i,t}$ using Equation~\ref{eqn5}. The diffusion model is trained using an MSE loss between the real and the predicted noise, summarized in Equation~\ref{eqn6}.} \label{fig2}
\end{figure}
The goal of this challenge is to replace tumor tissue with healthy tissue. We use the great performance of DDPMs in image synthesis to inpaint healthy tissue in masked regions of the brain. Figure~\ref{fig2} gives an overview.
\subsubsection{Training}
For the training process of the proposed approach, we need the ground truth image $x_{GT}$, a predefined mask $m$ that masks out some healthy tissue of the ground truth image, as well as the masked ground truth image denoted as $b$. Due to the high computation costs of 3D volumes, we propose a 2D approach on slices of the respective volumes. Therefore, we slice all 3D volumes into 2D slices, and consider only the slices $i$ with a non-zero mask for training. We take a random time step $t \in \{1,...,T\}$ and compute a noisy image $x_{i,t}$ from the ground truth slice $x_{i,GT}$ by applying Equation~\ref{eqn2} with $x_0$ = $x_{i,GT}$. The goal is to predict the slightly denoised $x_{i,t-1}$. We add the slice context information of our baseline image $b_i$ and the mask $m_i$ through channel-wise concatenation. The concatenated image $X_t:= x_{i,t} \oplus b_i \oplus m_i$ serves as input for our diffusion model. 
The computation of $x_{i,t-1}$ can be done via a denoising step with
\begin{equation}
x_{i,t-1}=\frac{1}{\sqrt[]{\alpha _{t}}}\left(x_{i,t}-\frac{1-\alpha _{t}}{\sqrt[]{1-\overline{\alpha }_{t}}}\epsilon _{\theta }(X_{t},t)\right)+\sigma_{t}\mathbf{z}, \quad \mbox{with } \mathbf{z} \sim \mathcal{N}(0,\mathbf{I}),
\label{eqn5}
\end{equation}
where $\epsilon _{\theta }(X_{t},t)$ is the diffusion model output at time step $t$, $\sigma_t$ is the variance scheme and $\mathbf{z}$ covers the stochastic component of the process. As shown in~\cite{durrer2023diffusion}, to train the diffusion model $\epsilon_{\theta}$ the following loss term with $\epsilon$ $\sim$ $\mathcal{N}(0,\mathbf{I})$ is used:
\begin{equation}
{\mid\mid\epsilon-\epsilon_{\theta }(X_{t},t)\mid\mid}^{2} = {\mid\mid\epsilon-\epsilon_{\theta }((\sqrt{\bar{a{_t}}}{x_{i,0}}+\sqrt{(1-\bar{a{_t}})}\epsilon)\oplus b_i \oplus m_i),t)\mid\mid}^{2}.
\label{eqn6}
\end{equation}

\newpage

\subsubsection{Sampling}
\begin{figure}
\centering 
\includegraphics[width=\textwidth]{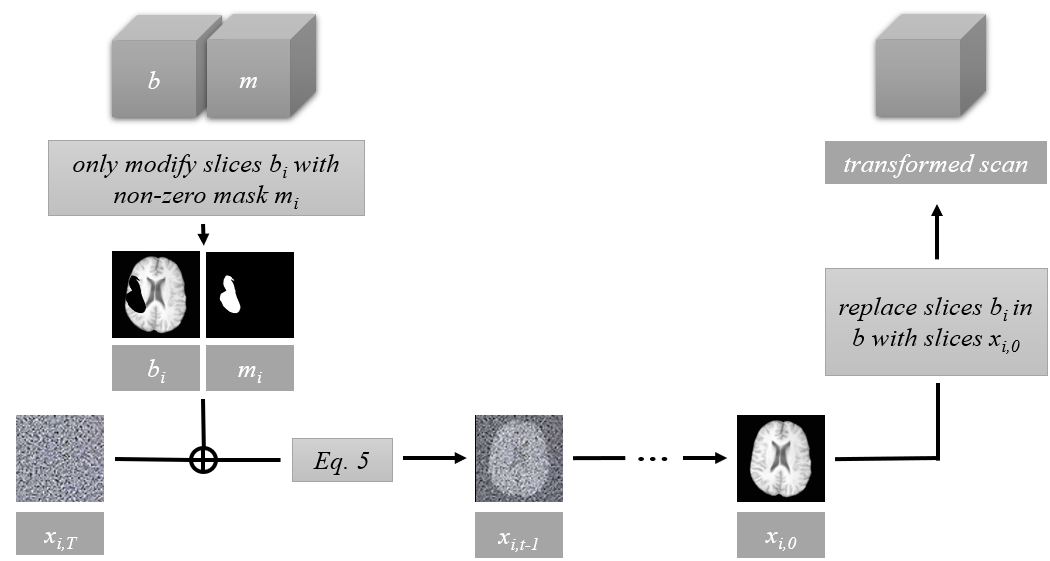}
\caption{Sampling method. During sampling, we apply the denoising described by Equation~\ref{eqn5} for each of the $T$ denoising steps for each slice $b_i$ with a non-zero mask $m_i$. We then replace the slices $b_i$ in $b$ with the samples $x_{i,0}$, containing the inpainting, to obtain our transformed 3D scan.} \label{fig3}
\end{figure}
During sampling, shown in Figure~\ref{fig3}, we loop through the slices of the masked 3D volume and only modify the slices with a non-zero mask. For each of these slices, the starting point $x_{i,T}$ is set to random Gaussian noise 
\newline $\sim$ $\mathcal{N}(0,\mathbf{I})$. We use the masked T1 slice $b_i$ and the corresponding mask $m_i$ to guide the denoising process for each time step $t$ for each of the slices. This means, we start from $x_{i,1000}$ $\sim$ $\mathcal{N}(0,\mathbf{I})$ and apply Equation~\ref{eqn5} until we get the prediction $x_{i,0}$ for each slice affected by a lesion. We replace the masked slices $b_i$ in the 3D scan $b$ with the predicted slices $x_{i,0}$ to get the transformed scan. This transformed scan can be used for further downstream tasks, as illustrated in Figure~\ref{fig1}.

\newpage

\section{Data Set and Training Details}
\subsection{Data  Preprocessing}
The data set consists of T1 scans from the multi-modal BraTS 2022 glioma segmentation challenge \cite{menze2014brats,bakas2017radiomic,bakas2018identifying,karargryris2023federated}. The data is a retrospective collection of brain tumor MR scans originating from multiple different institutions using different equipment and imaging protocols but all working under standard clinical conditions. For each patient, the challenge organizers provide T1 scans of brains affected by tumors, masks of the tumor regions, masks of healthy tissue regions, combined masks of the tumor and healthy regions as well as "T1-voided" scans, in which the combined masks are cropped out from the original T1 image. For our approach, we use the T1 images and the healthy region masks during training. As our final goal is to learn to replace healthy tissue, for training, we crop out the healthy masks from the T1 images to obtain a masked T1 image in which only a healthy region is missing.
We remove the top and bottom 0.1 percentile of voxel intensities and normalize the scans to voxel values between 0 and 1. As our model is constructed for 2D data, we slice each 3D scan into axial slices that are then cropped to a size of [224, 224]. Only background pixels are affected by the cropping. For training, we only use slices showing a non-zero mask. The provided training set consists of 1251 subjects which were used for a two-fold cross-validation, each using 1241 subjects for training and 10 for validation. 
\subsection{Training Details}
We trained our model on an NVIDIA A100-SXM4-40GB with batch size 8 for 2'850'000 iterations, taking about two and a half weeks. The number of channels in the first layer of the model is 128, as proposed in~\cite{wolleb2022diffusion}. Furthermore, one attention head is used at resolution 16, resulting in 113'672'066 model parameters. The learning rate is set to ${10^{-4}}$ for the used Adam optimizer. $T$ is defined as 1000. For the evaluation of the validation and test set, we use the model saved with an exponential moving average (EMA) over model parameters with a rate of 0.9999. Further details on hyperparameters and architecture can be found in~\cite{nichol2021improved}. 
\subsection{Postprocessing and Evaluation}
After sampling, we apply a Gaussian filter with a standard deviation $sigma = 1.075$ for the Gaussian kernel to smooth borders between the slices.
In addition to the training data, the challenge organizers provided a validation set of 219 subjects that could be evaluated using our model and submitted to the Synapse~\cite{synapse} platform to obtain structural similarity index measure (SSIM), peak signal-to-noise ratio (PSNR) and mean squared error (MSE). For submission to the Synapse platform, each model output is normalized to the intensity range of the corresponding input image from which the first and last 0.5 percentile were clipped. After submission of the final model as a MLCube~\cite{ml_cube}, it was evaluated by the challenge organizers on a non-public test set of 568 cases.

\section{Results}

\subsection{Cross-validation using Train-Validation Splits from Training Data Set}

We evaluated both models we trained for the cross-validation on our validation splits using the evaluation script provided by the organizers. The calculated metrics are SSIM, PSNR and MSE for the inpainted region only. Table~\ref{cv_test_split} provides the results of our cross-validation.

\begin{table}
\centering
\caption{Results of the two-fold cross validation.}\label{cv_test_split}
\begin{tabular}{c|c|c|c}

\bfseries & \bfseries{\textit{SSIM}} & \bfseries{\textit{PSNR}} & \bfseries{\textit{MSE}} \\
\hline
Model 1 & 0.8003 $[\pm 0.1081]$ & 18.7572 $[\pm 1.8850]$ & 0.0146 $[\pm 0.0063]$ \\
Model 2 & 0.7764 $[\pm 0.1292]$ & 18.7310 $[\pm 1.5019]$ & 0.0142 $[\pm 0.0049]$ \\
Average & \textbf{0.7884} $[\pm 0.1187]$ & \textbf{18.7441} $[\pm 1.6935]$ & \textbf{0.0144} $[\pm 0.0056]$ \\
\end{tabular}
\end{table}

\begin{figure}
\centering 
\includegraphics[width=0.9\textwidth]{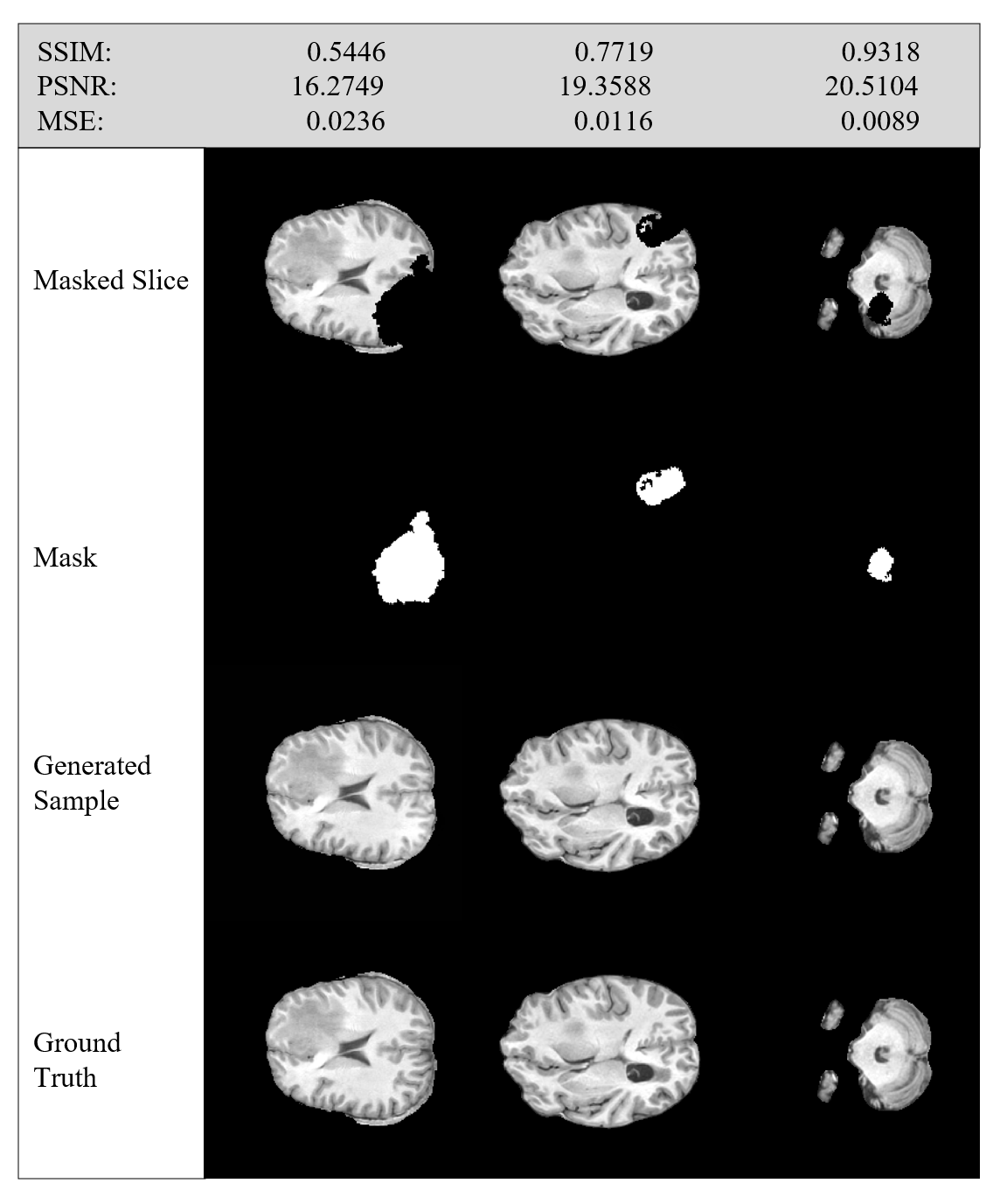}
\caption{Axial slices of images from our validation set. The masked slice, the mask, the generated sample, and the ground truth are provided. In general, the 2D generated samples are of good quality. Moreover, they show a high similarity to the corresponding ground truth. The SSIM, MSE and PSNR that are reported refer to the comparison of the whole generated volume (obtained by stacking of the generated slices) and the corresponding ground truth volume.} \label{fig4}
\end{figure}

\begin{figure}
\centering 
\includegraphics[width=0.8\textwidth]{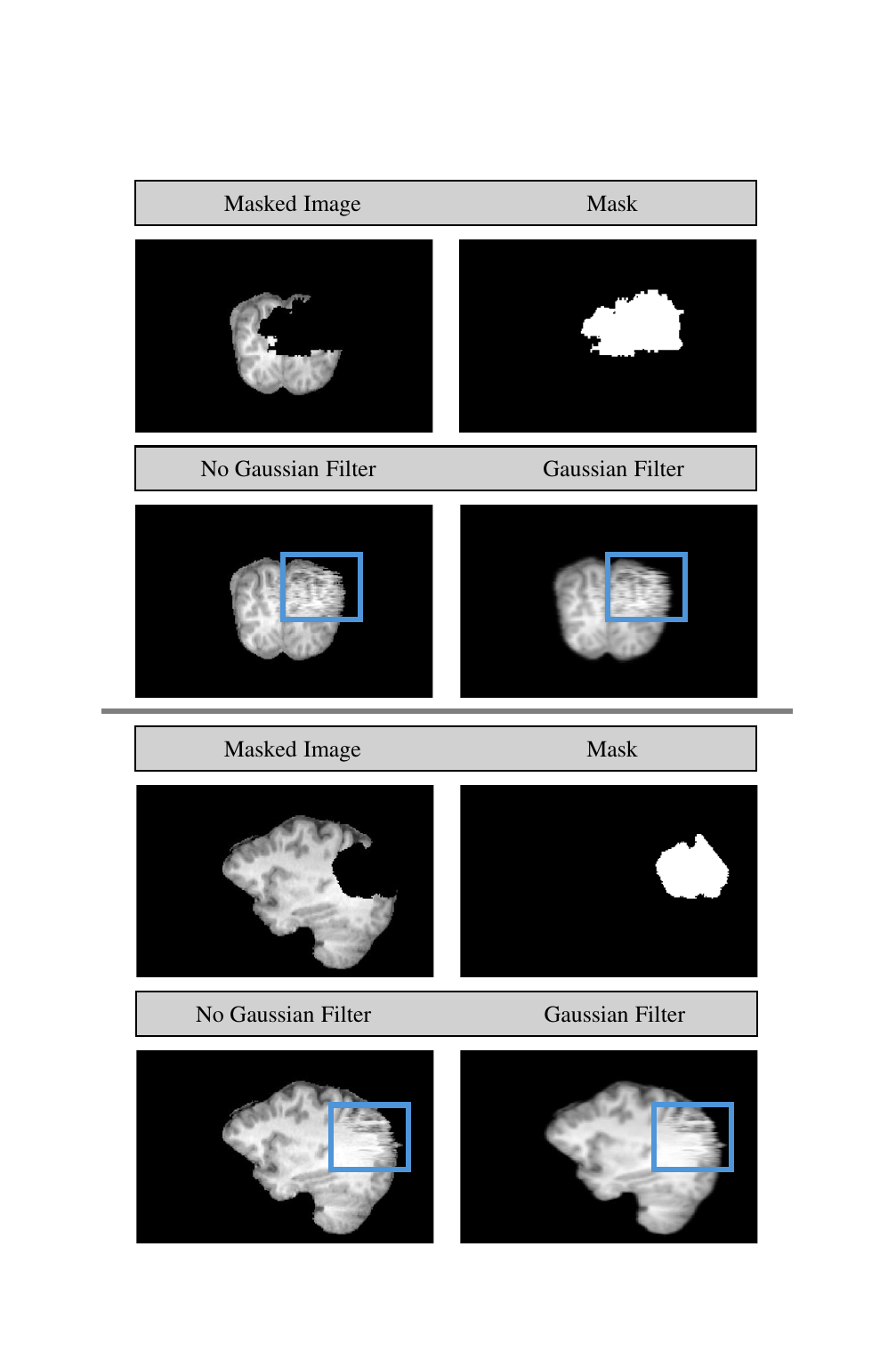}
\caption{Exemplary image of our validation set in the coronal (top) and sagittal (bottom) view. We observe stripe artifacts (blue boxes) due to the stacking of multiple axial slices. These artifacts can be reduced by applying a Gaussian filter (standard deviation $sigma = 1.075$ for Gaussian kernel).} \label{fig5}
\end{figure}

\noindent
Figure~\ref{fig4} provides examples of a given 2D masked slice, the corresponding mask, the generated sample and the ground truth.
In 2D, our approach manages to generate consistent results. However, if we look at the whole 3D volume, stripe artifacts appear in the sagittal and coronal plane if multiple sequential axial slices are replaced within the volume. This is shown in Figure~\ref{fig5}. One possibility to reduce the visibility of these artifacts is the application of a Gaussian filter, also demonstrated in Figure~\ref{fig5}.

\subsection{Validation on Synapse Server}

We evaluated both models that we trained for the cross-validation on the validation set provided by the organizers. The evaluation was performed on the Synapse server. Again, SSIM, PSNR and MSE are only evaluated for the inpainted region. Table~\ref{validation_scores} provides the results for our cross-validation. Regarding all metrics, we achieve comparable results as the other challenge participants.

\begin{table}[!h]
\centering
\caption{Results of the evaluation of our model's performance on the validation set, evaluated on the Synapse server.}\label{validation_scores}
\begin{tabular}{c|c|c|c}

\bfseries & \bfseries{\textit{SSIM}} & \bfseries{\textit{PSNR}} & \bfseries{\textit{MSE}} \\
\hline
Model 1 & 0.7802 $[\pm 0.1273]$ & 20.3220 $[\pm 2.8306]$ & 0.0113 $[\pm 0.0076]$ \\
Model 2 & 0.7805 $[\pm 0.1255]$ & 20.3830 $[\pm 2.7619]$ & 0.0112 $[\pm 0.0074]$ \\
Average & \textbf{0.7804} $[\pm 0.1264]$ & \textbf{20.3525} $[\pm 2.7963]$ & \textbf{0.0113} $[\pm 0.0075]$ \\

\end{tabular}
\end{table}

\subsection{Test Set Scores}

Model 2 was submitted as a MLCube for the evaluation on the non-public test set. The evaluation was performed by the challenge organizers. Table~\ref{test_scores} provides the obtained results. Again, SSIM, PSNR and MSE are only evaluated for the inpainted region. Using this model, we achieve the second place in the inpainting challenge.

\begin{table}[!h]
\centering
\caption{Results of the evaluation of our model on the non-public test set.}\label{test_scores}
\begin{tabular}{c|c|c|c}

\bfseries & \bfseries{\textit{SSIM}} & \bfseries{\textit{PSNR}} & \bfseries{\textit{MSE}} \\
\hline
Model 2 & 0.8271 $[\pm 0.1308]$ & 20.4949 $[\pm 3.1141]$ & 0.0115 $[\pm 0.0096]$ \\

\end{tabular}
\end{table}

\section{Discussion}
We present an approach for healthy tissue inpainting based on 2D DDPMs. Taking advantage of the excellent image synthesis performance of DDPMs, we apply these models to masked brain MR images. Image context of the masked image is provided by concatenating the masked image to the input of the diffusion model in every time step $t$. These concatenated images are created to guide the denoising process. As only healthy tissue is cropped out, the ground truth is available during training. Regarding the validation, we achieve similar results in terms of SSIM, PSNR and MSE as the other challenge participants. Our model creates good results in 2D, but a drawback of our method is that the stacking of the 2D slices can lead to stripe artifacts, which we tried to reduce using a Gaussian filter. As an alternative, we tried to avoid the stripe artifacts using a sequential approach, where we included the previous ground truth slice as an additional channel that was concatenated with the noisy image during training. During sampling, the previously sampled slice was concatenated with the noisy image. However, this approach still needs modifications, as it seemed as if the errors accumulated over the sequence of slices. An advantage of the model we used for this challenge is that due to the two-dimensionality, it needs less computational resources than 3D models and shows a stable training progression. Nevertheless, in future, we aim to expand our approach to a 3D model, e.g., similar to~\cite{bieder2023memory}. So far we experimented with a 3D model using the same framework as presented in this paper, but it was limited regarding memory consumption, which led to limitations in network architecture tuning. A drawback of diffusion models in general is the slow sampling speed compared to other generative models. We aim to improve this using e.g., Heun sampling schemes~\cite{karras2022heun} or consistency models~\cite{song2023consistency}. We conclude that our method provides a baseline that shows a good performance for the task, resulting in the second place in the competition, and has potential to be fine-tuned using different architectural and sampling changes.


%
%
%

\newpage

\begin{thebibliography}{8}


\bibitem{jenkinson2012fsl}
Jenkinson, M., Beckmann, C. F., Behrens, T. E., Woolrich, M. W., Smith, S. M.: Fsl. Neuroimage, \textbf{62}(2), pp. 182-790 (2012)

\bibitem{farazi2016inpainting}
Farazi, M. R., Faisal, F., Zaman, Z., Farhan, S.: Inpainting multiple sclerosis lesions for improving registration performance with brain atlas. In: International Conference on Medical Engineering, Health Informatics and Technology (2016)

\bibitem{almansour2021inpaint}
Almansour, M., Ghanem, N. M., Bassiouny, S.: High-Resolution MRI Brain Inpainting. In: IEEE EMBS International Conference on Biomedical and Health Informatics (2021)

\bibitem{menze2014brats}
Menze, B. H., Jakab, A., Bauer, S., Kalpathy-Cramer, J., Farahani, K., Kirby, J., Burren, Y.,Porz, N., Slotboom, J., Wiest, R., et al.: The multimodal brain tumor image segmentation benchmark (brats). IEEE transactions
on medical imaging \textbf{34}(10) 1993-2024 (2014)

\bibitem{bakas2017radiomic}
Bakas, S., Akbari, H., Sotiras, A., Bilello, M., Rozycki, M., Kirby, J. S., Freymann, J.B., Farahani, K., Davatzikos, C.: Advancing the cancer genome atlas glioma mri collections with expert segmentation labels
and radiomic features. Scientific data \textbf{4}(1) 1-13 (2017)

\bibitem{bakas2018identifying}
Bakas, S., Reyes, M., Jakab, A., Bauer, S., Rempfler, M., Crimi, A., Shinohara, R. T, Berger, C., Ha, S. M.,
Rozycki, M., et al.: Identifying the best machine learning algorithms for brain tumor segmentation, progression assessment, and overall survival prediction in the brats challenge. arXiv preprint arXiv:1811.02629 (2018)

\bibitem{baid2021rsna}
Baid, U., Ghodasara, S., Mohan, S., Bilello, M., Calabrese, E., Colak, E., Farahani, K., Kalpathy-Cramer, J., Kitamura, F. C., Pati, S., et al.: The rsna-asnr-miccai brats 2021 benchmark on brain tumor segmentation and radiogenomic classification. arXiv preprint arXiv:2107.02314 (2021)

\bibitem{kofler2023inpainting}
Kofler, F., et al.: The Brain Tumor Segmentation (BraTS) Challenge 2023: Local Synthesis of Healthy Brain Tissue via Inpainting. arXiv preprint arXiv:2305.08992 (2023)

\bibitem{karargryris2023federated}
Karargyris, A., Umeton, R., Sheller, M. J., Aristizabal, A., George, J., Wuest, A., Pati, S., et al.: Federated benchmarking of medical artificial intelligence with MedPerf. Nature Machine Intelligence, \textbf{5}, pp. 799-–810 (2023)

\bibitem{mamelak2007glioma}
Mamelak, A. N., Jacoby, D. B.: Targeted delivery of antitumoral therapy to glioma and other malignancies with synthetic chlorotoxin (TM-601). Expert Opinion on Drug Delivery, \textbf{4}(2), 175-182 (2007)

\bibitem{ho2020diffusion}
Ho, J., Jain, A., Abbeel, P.: Denoising diffusion probabilistic models. Advances in Neural Information Processing Systems, \textbf{33}, pp. 6840-6851 (2020)

\bibitem{saharia2022palette}
Saharia, C., Chan, W., Chang, H., Lee, C., Ho, J., Salimans, T, Fleet, D., Norouzi, M.: Palette: Image-to-image diffusion models. In: ACM SIGGRAPH 2022 Conference Proceedings, pp. 1-10 (2022)

\bibitem{wolleb2022diffusion}
Wolleb, J., Sandk{\"u}hler, R., Bieder, F., Valmaggia, P., Cattin, P.C: Diffusion Models for Implicit Image Segmentation Ensembles. In: Medical Imaging with Deep Learning, pp. 1336-1358. PMLR (2022)

\bibitem{durrer2023diffusion}
Durrer, A., Wolleb J., Bieder F., Sinnecker, T., Weigel M., Sandk{\"u}hler, R., Granziera, C., Yaldizli, {\"O}, Cattin, P.C: Diffusion Models for Contrast Harmonization of Magnetic Resonance Images. arXiv preprint, arXiv:2303.08189 (2023)

\bibitem{ronneberger2015unet}
Ronneberger, O., Fischer, P., Brox, T.: Convolutional networks for biomedical image segmentation. In: International Conference on Medical image computing and computer-assisted intervention, pp. 234-241. Springer (2021)

\bibitem{nichol2021improved}
Nichol, A. Q., Dhariwal, P.: Improved denoising diffusion probabilistic models. In: International Conference on Machine Learning, pp. 8162-8171. PMLR (2021)

\bibitem{synapse}
Synapse Homepage, \url{https://www.synapse.org/}. Last accessed 11 August 2023

\bibitem{ml_cube}
MLCube, \url{https://mlcommons.github.io/mlcube/}. Last accessed 11 August 2023

\bibitem{bieder2023memory}
Bieder, F., Wolleb, J., Durrer, A., Sandkuehler, R., Cattin, P.C.: Memory-Efficient 3D Denoising Diffusion Models for Medical Image Processing. In: Medical Imaging with Deep Learning (2023)

\bibitem{karras2022heun}
Karras, T., Aittala, M., Aila, T., Laine, S.: Elucidating the design space of diffusion-based generative models. Advances in Neural Information Processing Systems, \textbf{35}, pp. 26565-26577 (2022)

\bibitem{song2023consistency}
Song, Y., Dhariwal, P., Chen, M., Sutskever, I.: Consistency Models. arXiv preprint, arXiv:2303.01469 (2023)








\end{thebibliography}
%

\end{document}